\documentclass[twocolumn,aps,showpacs]{revtex4}
\usepackage{epsf,graphicx}
\usepackage{subfigure}
\def\Bf#1{\mbox{\boldmath $#1$}}
\begin{document}
\title{Dynamical Properties of a Rotating Bose-Einstein Condensate}
\author{Sebastian Kling}
\email{kling@iap.uni-bonn.de}
\affiliation{Institut f{\"u}r Angewandte Physik, Universit{\"a}t Bonn, Wegelerstra{\ss}e 8,
53115 Bonn, Germany}
\author{Axel Pelster}
\email{axel.pelster@uni-due.de}
\affiliation{Fachbereich Physik, Campus Duisburg, Universit{\"a}t Duisburg-Essen, 
Lotharstra{\ss}e 1, 47048 Duisburg, Germany}      

\date{\today}

\begin{abstract}
Within a variational approach to solve the Gross-Pitaevskii equation we investigate dynamical properties
of a rotating Bose-Einstein condensate which is confined in an anharmonic trap. In particular, we
calculate the eigenfrequencies of low-energy excitations out of the equilibrium state and the
aspect ratio of the condensate widths during the free expansion.
\end{abstract}
\maketitle
\section{Introduction}
The rotation of a Bose-Einstein condensate (BEC) has raised enormous interest, both experimentally 
and theoretically over the last few years \cite{Review}. 
The BEC is a superfluid, hence its rotational behavior is quite different from that of a normal fluid. 
Its properties strongly depend on the specific trap in which it is confined.
In harmonic rotating traps centrifugal forces may break the confinement when the rotation frequency gets close to the
trapping frequency.
To overcome this centrifugal barrier, one has to increase the strength of the radial confinement. 
The most natural approach relies on adding a radial quartic potential to the harmonic trap. 
Such an anharmonic trap was realized in a recent experiment in Paris at the {\'E}cole Normale Sup{\'e}rieure 
(ENS) in the group of J.~Dalibard, where the centrifugal limit can be exceeded slightly \cite{Bretin,Stock06}.
This combined trap changes both the thermodynamical and the dynamical properties of the condensate 
drastically, in particular in the fast rotation regime. Nevertheless, 
the analysis of the experimental data has been mainly performed by considering only a harmonic trap. 
This approximation is valid for a slow rotation frequency, but a correct 
interpretation of the experimental data in the fast rotation regime necessitates an analysis for the full
anharmonic trap. Whereas the proceeding paper \cite{Kling1} elaborates 
within the semiclassical approximation the thermodynamical properties
of such a rotating BEC as, for instance, the critical temperature and the heat capacity, the present
work investigates the corresponding dynamical properties. To this end we proceed as follows.
Section \ref{HF} introduces the basic Gross-Pitaevskii equation for the condensate wave function of a 
rotating trapped BEC. Neglecting the emergence of vortices, we determine in  Section \ref{SP}
the stationary equilibrium density of a rotating condensate within the Thomas-Fermi approximation. 
In Section \ref{DP} we then use a variational approach to investigate the hydrodynamic modes of the condensate and its 
free expansion after switching off the trap. 
All results are obtained for the specific parameters of the Paris experiment \cite{Bretin,Stock06}.
\section{Hydrodynamic Formalism}\label{HF}
At zero temperature, the dynamics of a BEC is described by the Gross-Pitaevskii Lagrangian density
\cite{Gross61,Pita61}
\begin{eqnarray}
\label{GPL1}
\mathcal{L}&=&i\hbar\psi^*\frac{\partial}{\partial t}\psi-\frac{\hbar^2}{2M}{\Bf \nabla}
\psi^* \cdot {\Bf \nabla}\psi\nonumber\\[2mm]
&&-V({\bf x})\psi^*\psi-\frac{2\pi\hbar^2a_s}{M}(\psi^*\psi)^2\,.
\end{eqnarray}
Here, $V({\bf x})$ denotes the trapping potential, $a_s$ the s-wave scattering length, and $M$ the atomic mass.
The transformation into the rotating frame has the consequence that an angular momentum $-{\bf \Omega}{\bf 
\hat{\hspace*{0.7mm}L}}$ is added to the Lagrangian (\ref{GPL1}), 
where the rotation vector is denoted by ${\bf \Omega}=\Omega {\bf e}_z$ and ${\bf \hat{\hspace*{0.7mm}L}}$
represents the one-particle 
quantum mechanical angular momentum operator \cite{Pethick,Pitaevskii}.
Thus, the Euler-Lagrange equation yields the Gross-Pitaevskii (GP) equation 
\begin{equation}
\label{GP1}
i\hbar\frac{\partial}{\partial t}\psi \!=\! \left[-\frac{\hbar^2}{2M}\Delta + V({\bf x}) 
+\frac{4\pi\hbar^2a_s}{M}|\psi|^2+{\bf \Omega {\bf 
\hat{\hspace*{0.7mm}L}}} \right]\psi \,,
\end{equation}
Performing the decomposition $\psi=|\psi|e^{iS}$  we work out the hydrodynamic aspects of the system, with condensate 
density $n=|\psi|^2$ and superfluid velocity ${\bf v}_{\rm s}=\hbar {\Bf \nabla} S / M$. 
Basic manipulations yield for them the coupled
hydrodynamic equations which consist of a continuity equation  
\begin{equation}
\label{GPF3}
\frac{\partial}{\partial t}n = -{\Bf \nabla}\left[({\bf v}_{\rm s}-{\bf v}_{\rm sb})n\right]
\end{equation}
and the Euler equation 
\begin{eqnarray}
\label{GPF2}
M \frac{\partial}{\partial t}{\bf v}_{\rm s} &=& - {\Bf \nabla}\left[
\frac{M}{2}({\bf v}_{\rm s}-{\bf v}_{\rm sb})^2
- \frac{\hbar^2}{2M} \frac{\Delta \sqrt{n}}{ \sqrt{n}} \right.\nonumber\\[2mm]
&&\left. + V({\bf x},\Omega)+ \frac{4\pi\hbar^2a_s}{M}n \right]\,,
\end{eqnarray}
where $V({\bf x}, \Omega )= V({\bf x})- M \Omega^2 (x^2 + y^2 ) / 2$ denotes the effective potential
in the rotating frame.
Here, the solid-body velocity is denoted by ${\bf v}_{\rm sb}={\bf x} \times {\bf \Omega}$. 
In the following, we consider the cylindrical anharmonic trap used in the Paris experiment \cite{Bretin,Stock06}
\begin{equation}
\label{Vrot1}
V({\bf x},\Omega)=\frac{M}{2}\left(\omega_\bot^2-\Omega^2\right)r_\bot^2+\frac{M}{2}\omega_z^2z^2+\frac{k}{4}r_\bot^4\,,
\end{equation}
where $M$ is the atomic mass of ${}^{87}$Rb, $\omega_\bot=2\pi\cdot64.8$~Hz, $\omega_z= 2\pi\cdot11.0$~Hz, 
$k=2.6\cdot10^{-11}$~Jm${}^{-4}$, 
and $r_\bot=\sqrt{x^2+y^2}$.
An amount of $3\cdot 10^{5}$ atoms has been condensed and the  highest realized rotation speed was
$\Omega_{\rm max}\approx 1.04\cdot\omega_\bot$. \\
The superfluid is irrotational ${\Bf \nabla}\times{\bf v}_{\rm s}={\bf 0}$, except at singularities, whereas 
${\Bf \nabla}\times{\bf v}_{\rm sb}=- 2{\bf \Omega}$.
Thus the difference of both ${\bf v}_{\rm s}-{\bf v}_{\rm sb}$ cannot vanish identically. 
However, the superfluid can mimic solid-body rotation by forming an uniform array of rectilinear vortices parallel 
to the axis of rotation \cite{Feyn55}, which minimizes the difference ${\bf v}_{\rm s}-{\bf v}_{\rm sb}$.
In the detailed analysis of Ref.~\cite{Tka66} it has been shown that the triangular lattice is the energetically most 
favorable one. 
For a dense vortex lattice it is even justified to neglect in the Euler equation (\ref{GPF2})
the term proportional to the difference 
${\bf v}_{\rm s}-{\bf v}_{\rm sb}$ \cite{Fetter83}. In the following we assume that this approximation is valid
and focus on investigating the effect of the centrifugal term in the potential (\ref{Vrot1}).
\setlength{\unitlength}{1cm}
\begin{figure}[t]
\center\begin{minipage}[tb]{.4\textwidth}
\center\includegraphics[scale=1.0]{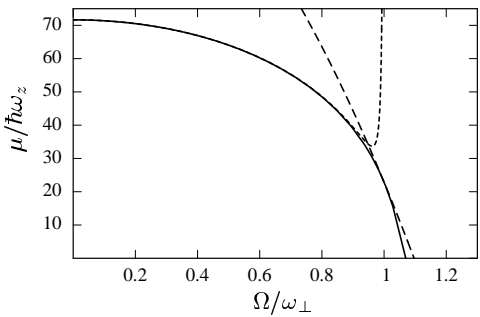}
\caption{\small Chemical potential versus rotation frequency for the values of the Paris experiment
\cite{Bretin,Stock06}. The solid line corresponds to (\ref{NC4}), the short-dashed and the long-dashed line
to the approximation (\ref{Mu2}) and (\ref{Mu3}), respectively.}
\label{CM}
\end{minipage}
\end{figure}
\section{Static Properties}\label{SP}
\begin{figure}[t]
\center\begin{minipage}[tb]{.4\textwidth}
\center\includegraphics[scale=1.0]{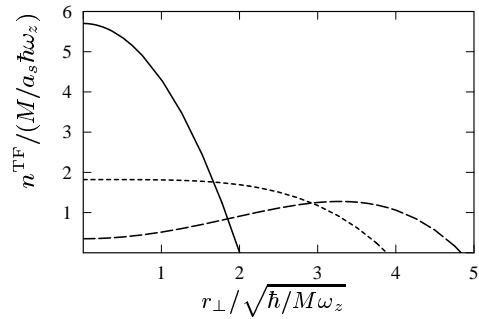}
\caption{\small Thomas-Fermi density (\ref{TFD1}) in the $xy$-plane for the values of the Paris experiment 
\cite{Bretin,Stock06} and various rotation frequencies $\Omega$. From top to bottom the rotation frequencies 
are $\Omega/\omega_\bot=0$ (solid), $\Omega/\omega_\bot=1$ (short-dashed) and $\Omega/\omega_\bot=1.06$ (long-dashed).}
\label{TFD}
\end{minipage}
\end{figure}

We start with analyzing the static properties of such a condensate, where we follow Ref.~\cite{Fetter01}. 
To this end we separate the time-dependence according to $n=n({\bf x})$ and $S=-\mu t/ \hbar + s({\bf x})$, 
where $\mu$ denotes the chemical potential.
With this we obtain from the Euler equation (\ref{GPF2}) the stationary hydrodynamic equation
\begin{eqnarray}
\label{GPS1}
\mu &=& - \frac{\hbar^2}{2Mn} \frac{\Delta \sqrt{n}}{ \sqrt{n}} +V({\bf x},\Omega) + \frac{4\pi\hbar^2a_s}{M}n\,.
\end{eqnarray}
Furthermore, we perform the Thomas-Fermi (TF) approximation \cite{Baym96}, i.e. the 
kinetic energy is neglected in (\ref{GPS1}), so that the condensate density reads
\begin{equation}
\label{TFD1}
n^{\rm TF}\approx\frac{M}{4\pi\hbar^2a_s}\big[\mu-V({\bf x},\Omega)\big]\Theta\big[\mu-V({\bf x},\Omega)\big]\,,
\end{equation}
where the Heaviside function $\Theta$ ensures its positivity.
The chemical potential is determined by the normalization condition $N=\int d^3x\,n^{\rm TF}$. 
The Thomas-Fermi radii for the cylinder symmetric trap potential (\ref{Vrot1}) read
\begin{eqnarray}
\label{TFR1}
\hspace*{-5mm}R_z^{\rm TF} &=& \sqrt{\frac{2 \mu}{M \omega_z^2}} \, , \\
\label{TFR2}
\hspace*{-5mm}R_\bot^{\rm TF}& =& \sqrt{ \frac{M(\Omega^2- \omega_\bot^2)}{k} + \sqrt{ 
\frac{M^2(\Omega^2-\omega_\bot^2)^2}{k^2} + \frac{4\mu}
{k} } } \, .
\end{eqnarray}
We work out the spatial integration in cylinder coordinates, yielding
\begin{eqnarray}
\label{NC4}
N &=& \frac{2\sqrt{2M}\mu^2}{3a_s\sqrt{k}\hbar^2\omega_z}f\left(\frac{M(\omega_\bot^2-\Omega^2)}{2\sqrt{k\mu}}\right)\,,
\end{eqnarray}
where the function 
\begin{eqnarray}
\label{f1}
f(x)&=&\frac{3\pi}{16} \left( 1+x^2 \right)^2 \left[ 1-\frac{2}{\pi}\arcsin{\frac{x}{\sqrt{1+x^2}\,}} \right] 
\nonumber\\[2mm]
&&- \frac{5x}{8} - \frac{3x^3}{8}
\end{eqnarray}
has the asymptotic properties
%
%
\setlength{\unitlength}{1cm}
\begin{figure}[t]
\center\begin{minipage}[tb]{.4\textwidth}
\center\includegraphics[scale=1.0]{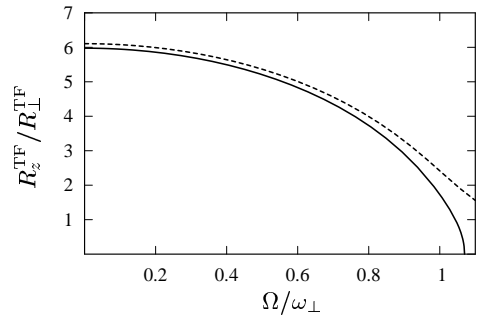}
\caption{\small Aspect ratio of condensate radii versus rotation frequency for the values of the Paris 
experiment \cite{Bretin,Stock06}. Solid line corresponds to the Thomas-Fermi radii 
$R_z^{\rm TF}/R_{\bot}^{\rm TF}$, dashed 
line to the stationary radii $W_{0z}^{\rm TF}/W_{0x}^{\rm TF}$ determined from Eq.~(\ref{RESU}) in Section~\ref{LLO}.}
\label{TFR}
\end{minipage}
\end{figure}
\begin{equation}
\label{f2}
f(x) =\left\{\begin{array}{lll}x/5  -x^3/35&{\rm for}&x\gg 1\\[2mm]
3\pi(x^2+1)^2/16-x-x^3&{\rm for}&|x|\ll 1\\[2mm]
3\pi(x^2+1)^2/8-x/5&{\rm for}&x\ll -1\,.
\end{array}\right.
\end{equation}
When the rotation frequency $\Omega$ varies between $0$ and $\Omega_{\rm max}= 1.04\cdot\omega_\bot$, the argument 
$x=M(\omega_\bot^2-\Omega^2)/(2\sqrt{k\mu})$ of the 
function (\ref{f1}) ranges between $-\infty<x<3.3$. 
A slow rotation  corresponds to $x\gg1$, where the chemical potential reads
\begin{eqnarray}
\label{Mu2}
\hspace*{-0.5cm} \mu &\approx& \hbar\omega_z
\left[ \frac{15a_sN\sqrt{M}(\omega_\bot^2-\Omega^2)}{4\sqrt{2\hbar\omega_z^3}} \right]^{2/5}\nonumber\\[2mm]
\hspace*{-0.5cm}&&\times\left\{1 + \frac{4k\left[ 15a_sN\sqrt{M}(\omega_\bot^2-\Omega^2)\right]^{5/2}}{35M^2(\omega_\bot^2
-\Omega^2)^2\sqrt{\hbar\omega_z^3}}\right\}\,. 
\end{eqnarray}
For the fast rotation regime $\Omega\sim\omega_\bot$ we have to use the second expansion (\ref{f2}), where $|x|\ll1$, so 
we obtain for the chemical potential 
\begin{eqnarray}
\label{Mu3}
\hspace*{-0.5cm}\mu &\approx&\hbar\omega_z
\left(\frac{32N^2a_s^2 k}{\pi^2 M\omega_z^2}\right)^{1/4}\nonumber\\[2mm]
\hspace*{-0.5cm}&& \times
\left[1 + \frac{4}{3\pi}\left(\frac{32N^2a_s^2 k}{\pi^2 M\omega_z^2}\right)^{1/8}\frac{M(\omega_\bot^2
-\Omega^2)}{\sqrt{\hbar\omega_z k}}\right]\,.
\end{eqnarray}
For an arbitrary rotation frequency $\Omega$ the chemical potential follows directly from numerically inverting 
(\ref{NC4}). According to Figure \ref{CM} we observe that the chemical potential  vanishes 
for the experimental data of the Paris experiment 
\cite{Bretin,Stock06} at a critical 
rotation frequency $\Omega_c \approx1.07\cdot\omega_\bot$. Our assumption, that this indicates the emergence
of an instability, is confirmed by evaluating the TF radii (\ref{TFR1}), (\ref{TFR2}) and the TF density (\ref{TFD1}).
Figure \ref{TFR} depicts that the TF radius in $z$-direction vanishes 
with the chemical potential as it is proportional to the square root of it, 
while the other radius in the perpendicular plane remains finite. Note that the vanishing of one TF radius also
signals the breakdown of the validity of the TF approximation. Thus, our TF value  $\Omega_c \approx1.07\cdot\omega_\bot$
for the critical rotation frequency should be modified by a full quantum mechanical calculation.
In Figure~\ref{TFD} the TF density is shown for three particular rotation frequencies. 
For an overcritical rotation $\Omega>\omega_\bot$, we notice a pronounced dell in the density profile. 
Thus, we can conclude from Figures~\ref{CM}--\ref{TFD} that our TF analysis for the stationary condensate
density qualitatively agrees with the experimental finding that the rotation frequency can only be increased until
$\Omega_{\rm max}\approx 1.04\cdot\omega_\bot$.
\setlength{\unitlength}{1cm}
\section{Dynamic Properties}\label{DP}
In order to investigate the dynamical properties of the condensate, we return to the Euler equation (\ref{GPF2})  
in the TF approximation and consider collective excitations of the stationary state. 
Thus, we look for solutions of the type $n ({\bf x},t)= n^{\rm TF}({\bf x})+\delta n({\bf x})\,e^{i\omega t}$, 
where $n^{\rm TF}({\bf x})$ is the stationary TF 
density (\ref{TFD1}) and $\delta n({\bf x})\,e^{i\omega t}$ a small time-dependent deviation. 
Analogously, the superfluid
velocity field is decomposed according to ${\bf v}_{\rm s}= {\bf v}_{\rm sb}+\delta {\bf v}$. 
In first order of $\delta n$ and $\delta {\bf v}$ we obtain the eigenmode equation by taking the time derivative of 
the Euler equation (\ref{GPF2}) and inserting the continuity equation (\ref{GPF3}) 
\begin{equation}
\label{EM1}
M\omega^2 \delta n = -\left[\mu - V({\bf x},\Omega)\right]{\Bf \nabla}^2\delta n + {\Bf \nabla} V({\bf x},\Omega)
\cdot{\Bf \nabla}\delta n\,,
\end{equation}
where the chemical potential is a constant which is determined from normalizing the stationary density as 
described in the previous section. 
The eigenmode equation (\ref{EM1}) has been solved analytically for a harmonic trap which is
isotropic \cite{Str96} or rotationally symmetric \cite{Fl97}. 
However, investigating Eq. (\ref{EM1}) for the anharmonic trap, we observe that the differential 
operator does not conserve powers of the 
radial coordinate $r_\bot$, so that we may not expect simple analytical solutions. 
In the special case of an overcritical rotation 
$\Omega>\omega_\bot$ an analytical solution of (14) was found in Ref.~\cite{Fetter05}. \\
Another way to obtain the eigenmodes at least approximatively
is a variational approach which is based on Ritz's method. 
This method was already successfully applied for a harmonic trap \cite{PG96} and is in agreement with the solution of 
(\ref{EM1}) for a large two-particle $\delta$-interaction, i.e.~the regime where the TF approximation is valid. 
For a simplified one- and two-dimensional case 
this method was already applied for the anharmonic trap (\ref{Vrot1}) \cite{Li06,Ghosh04}. 
However, a correct description involves a full three dimensional treatment which we present now. 
\subsection{Variational Method}\label{varmet}
For our variational approach we follow Ref.~\cite{PG96} and use a Gaussian shape of the condensate
%
\setlength{\unitlength}{1cm}
\begin{figure}[t]
\center\begin{minipage}[Ct]{.4\textwidth}
\includegraphics[scale=1.0]{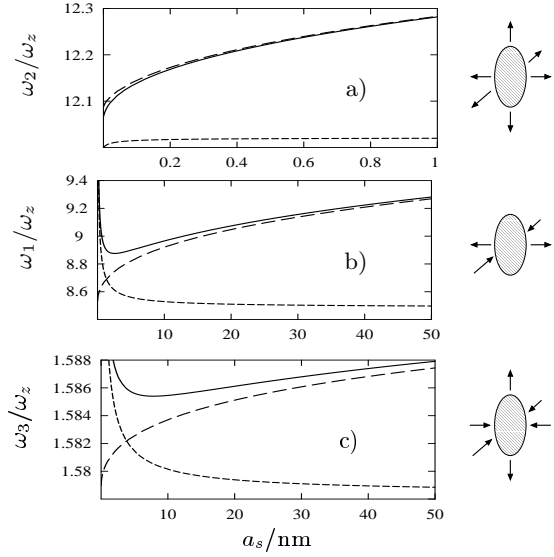}
\caption{\small Three eigenmodes of the non-rotating condensate ($\Omega=0$) confined 
in the anharmonic trap (\ref{Vrot1}):
a) breathing mode, b) and c) scissor modes.
The solid lines correspond to the solution of the variational procedure (\ref{EM3}), the long-dashed 
lines to its TF approximation (\ref{EM4}), and the short-dashed lines to the harmonic trap ($\kappa=0$).
The condensate clouds visualize the temporal behaviour
of the respective eigenmodes.}
\label{HDM1}
\end{minipage}
\end{figure}
%
\begin{eqnarray}
\psi({\bf x},t) &=& \frac{1}{\sqrt{\pi^{3/2}W_x W_y W_z}}
\exp{\left\{ \frac{x^2}{2W_x^2}+\frac{y^2}{2W_y^2}\right.}
\nonumber\\
&&\left.+\frac{z^2}{2W_z^2} +i\left[S_x x^2+S_y y^2+S_z z^2\right]\right\} \,,
\label{GT1}
\end{eqnarray}
where $W_{x,y,z}$ and $S_{x,y,z}$ denote 
time-dependent variational parameters for the widths and the phases, respectively. 
Note that (\ref{GT1}) can only be used in the undercritical regime $\Omega\leq\omega_\bot$. For $\Omega>\omega_\bot$
a central hole emerges in the condensate (see Figure \ref{TFD}), which is not described by 
the Gaussian trial function. For the ansatz
(\ref{GT1}) we determine the Lagrangian $L=\int d^3x\,\mathcal{L}$ with the density (\ref{GPL1})
and the potential (\ref{Vrot1}). 
Thus, the resulting action $A=\int dt\, L$ represents a functional of the trial functions $W_{x,y,z}$ 
and $S_{x,y,z}$ which are determined according to the Hamilton principle. 
As we find for the phases the relation $S_{x,y,z}=-\dot{W}_{x,y,z}/ 2W_{x,y,z}$, we can eliminate them
and find for the widths the coupled equations of motion
\begin{eqnarray}
\label{EM2}
\frac{d^2}{d t^2}W_{x,y}&=&-\lambda^2\eta W_{x,y} + \frac{1}{W_{x,y}^3}+\frac{P}{W_{x,y}^2W_{y,x}W_z}\nonumber\\
&&-\kappa\left( 3W_{x,y}^3+W_{x,y}W_{y,x}^2 \right)\,,\nonumber\\[2mm]
\frac{d^2}{d t^2}W_z&=&-W_z + \frac{1}{W_z^3}+\frac{P}{W_xW_yW_z^2} \,,
\end{eqnarray}
where we have introduced the dimensionless variables $\lambda=\omega_\bot/\omega_z\approx 6$, 
$\eta=1-(\Omega/\omega_\bot)^2$, and $\kappa=k\hbar\omega_z/a_z^4\approx 0.4$.
The length scale is set by the harmonic oscillator length $a_z=\sqrt{\hbar/M\omega_z}$, so that 
$W_{x,y,z}\to W_{x,y,z} a_z$, and the time is scaled by $t\to t/\omega_z$. 
The resulting dimensionless interaction parameter is denoted by $P=\sqrt{2/\pi}Na_s/a_z$.
The term proportional to $\kappa$ is due to the anharmonicity of the trap and represents the main difference
in comparison with the treatment of the harmonic trap in Ref.~\cite{PG96}.
The Euler-Lagrange equations (\ref{EM2}) are of the form $\ddot{W}_{x,y,z} = 
-\partial V_{\mathrm{eff}}(W_x,W_y,W_z)/\partial \,W_{x,y,z}$ and can therefore be regarded as the motion of 
a fictitious point particle in the effective potential
\begin{eqnarray}
\label{Veff1}
V_{\mathrm{eff}}({\bf W})&&=\frac{\lambda^2\eta}{2}W_x^2+\frac{\lambda^2\eta}{2}W_y^2+\frac{1}{2W_x^2}
+\frac{1}{2W_y^2}\nonumber\\
&&+\frac{\kappa}{4}(3W_x^4+2W_x^2W_y^2+3W_y^4) \nonumber \\ 
&&+\frac{1}{2}W_z^2+\frac{1}{2W_z^2}+\frac{P}{W_xW_yW_z}\,,
\end{eqnarray}
where the vector of the width is denoted by ${\bf W}=(W_x,W_y,W_z)$.
\setlength{\unitlength}{1cm}
\begin{figure}[t]
\center\begin{minipage}[t]{.4\textwidth}
\center\includegraphics[scale=1.0]{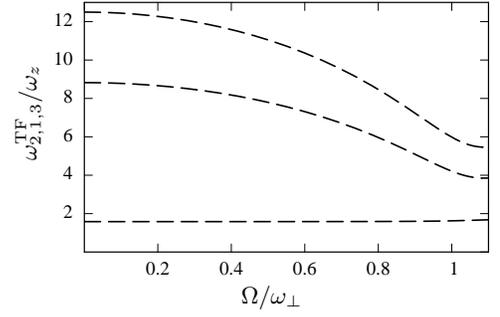}
\caption{\small Eigenfrequencies of the condensate confined in the anharmonic trap (\ref{Vrot1}) in the
TF approximation (\ref{EM4}).}
\label{HDM2}
\end{minipage}
\end{figure}
%
\subsection{Low-Energy Excitations}\label{LLO}
At first, we determine the optimal widths of the equilibrium state around which the density fluctuates. 
The stationary point is obtained by setting $\ddot{W}_{0k}=0$, yielding a set of algebraic equations. 
Because of the rotational symmetry of the trap we have $W_{0x}=W_{0y}$, thus the stationary widths follow from
\begin{eqnarray}
\lambda^2 \eta W_{0x}&=& \frac{1}{W_{0x}^3} + \frac{P}{W_{0x}^3 W_{0z}} - 4 \kappa W_{0x}^3 \, , \nonumber\\
W_{0z} & = & \frac{1}{W_{0z}^3}+ \frac{P}{W_{0x}^2 W_{0z}^2} \, . \label{SW}
\end{eqnarray}
Considering small deviations of the equilibrium state, we perform a Taylor approximation of the effective potential 
$V_{\mathrm{eff}}\,({\bf W}) \approx V_{\mathrm{eff}}\,({\bf W}_0)+\delta {\bf W}\,H\,\delta {\bf W}^T / 2$, 
where ${\bf W}_0$ 
denotes the stationary point and $H$ is the Hessian matrix. 
The square-root of the eigenvalues of $H$ are the low excitation frequencies, for which we find the analytical expressions
\begin{eqnarray}
\omega_1&=&2 \omega_z\,\sqrt{\lambda^2\eta +5\kappa W_{0x}^2-2P_{4,1}},\nonumber\\[2mm]
\omega_{2,3}&=&\sqrt{2}\omega_z\, \Bigg(1+\lambda^2\eta +6\kappa W_{0x}^2-P_{2,3}\Bigg.\label{EM3}\\[1mm]
&&\hspace{-12mm}\Bigg.\pm\sqrt{\big[\lambda^2\eta+6\kappa W_{0x}^2-1+P_{2,3}\big]^2
+8P_{2,3} P_{4,1}}\,\Bigg)^{1/2}\!,\nonumber
\end{eqnarray}
%
%
\setlength{\unitlength}{1cm}
\begin{figure}[t]
\center\begin{minipage}[t]{.4\textwidth}
\center\includegraphics[scale=0.78]{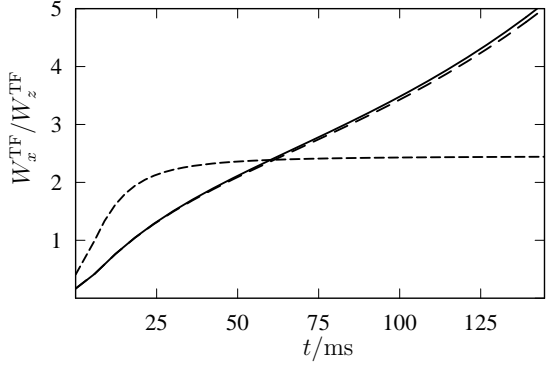}
\caption{\small Aspect ratio of the non-rotating condensate ($\Omega=0$) for the values of the Paris experiment 
\cite{Bretin,Stock06}. The solid line corresponds to $\kappa=0.4$, the long-dashed lines to $\kappa=0$, and the
short-dashed lines to the interaction free Bose gas ($a_s=0$) with $\kappa=0.4$.}
\label{Ex1}
\end{minipage}
\end{figure}
\hspace*{-3.5mm}where we have introduced the notation $P_{i,j}=P/(4W_{0x}^iW_{0z}^j)$.
The corresponding TF approximation of these modes follows from neglecting the 
$1/W^3$-terms in (\ref{EM2}) which 
correspond to the kinetic energy and is here related to a large interaction parameter $P\to\infty$. 
Thus, the algebraic equations (\ref{SW}) for the stationary widths reduce to
\begin{eqnarray}
\label{RESU}
\hspace*{-2mm}P=W_{0x}^{\rm TF\,2}W_{0z}^{\rm TF\,3}\, , \, W_{0z}^{\rm TF\,2}=W_{0x}^{\rm TF\,2}(\lambda^2\eta
+4\kappa W_{0x}^{\rm TF\,2}) \, . 
\end{eqnarray}
The resulting ratio $ W_{0z}^{\rm TF} / W_{0x}^{\rm TF}$ 
agrees qualitatively with the aspect ratio of TF condensate radii (see Figure \ref{TFR}).
The oscillation frequencies (\ref{EM3}) simplify in the TF approximation to
\begin{eqnarray}
\label{EM4}
\omega_{1}^{\rm TF}&=&\sqrt{2}\omega_z\,\sqrt{\lambda^2\eta + 6\kappa W_{0x}^{\rm TF\,2}} \,,\label{e1tf}\nonumber\\[2mm]
\omega_{2,3}^{\rm TF}&=&\sqrt{2} \omega_z\Bigg( 3/4 + \lambda^2\eta + 6\kappa W_{0x}^{\rm TF\,2}\Bigg.\\[1mm]
&&\hspace{-1.5cm}\Bigg.\pm\sqrt{\left(\lambda^2\eta + 6\kappa W_{0x}^{\rm TF\,2} - 3/4 \right)^2 
+2 \kappa W_{0x}^{\rm TF\,2} 
+ \lambda^2\eta /2} \,\Bigg)^{1/2}\hspace*{-2mm}.\nonumber
\end{eqnarray}
Figure \ref{HDM1} shows that the three eigenfrequencies (\ref{EM3}) and their TF approximation (\ref{EM4}) for 
a non-rotating condensate ($\Omega=0$) increase with the s-wave scattering length $a_s$. 
This effect is due to the presence of the trap anharmonicity, as the eigenfrequencies in a harmonic trap ($\kappa=0$)
tend to a constant in the limit $a_s \to \infty$.
For ${}^{87}$Rb the s-wave scattering length is about $a_s=5$~nm, so we can conclude 
from Figure \ref{HDM1} that the TF approximation is 
sufficient for the Paris experiment \cite{Bretin,Stock06}. Therefore, the resulting strong
dependence of the eigenmodes on the rotation frequency $\Omega$ in Figure \ref{HDM2} is only depicted 
for the TF approximated frequencies (\ref{EM4}). 
%
\setlength{\unitlength}{1cm}
\begin{figure}[t]
\center\begin{minipage}[t]{.4\textwidth}
\center\includegraphics[scale=0.8]{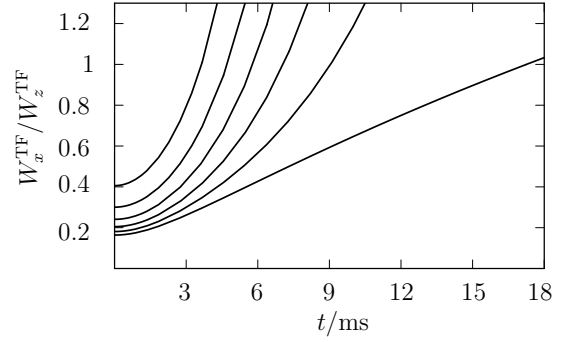}
\caption{\small Aspect ratio of the rotating condensate confined in the 
Paris trap \cite{Bretin,Stock06} for various rotation frequencies. 
The rotation parameter $\eta=1-(\Omega/\omega_\bot)^2$ is varied in steps of width $0.2$ from $0$ 
(lowest line) to $1$ (highest line).}
\label{Ex2}
\end{minipage}
\end{figure}
\subsection{Free Expansion}
The equations of motion (\ref{EM2}) determine also the expansion dynamics of the condensate after switching off the trap. 
Following Ref.~\cite{CasDum96}, we rewrite them within the TF approximation, yielding 
\begin{eqnarray}
\label{EM5}
\ddot{b}_\bot &=& \frac{\Omega^2b_\bot}{\omega_z^2} + 
\frac{\lambda^2\eta + 4\kappa W_{0x}^{\rm TF\,2}}{b_\bot^3b_z}\,,\\
\ddot{b}_z &=&\frac{1}{b_\bot^2b_z^2}\,, \label{EM5B}
\end{eqnarray}
where we have introduced the scaling parameters $b_\bot(t)=W_{x}^{\rm TF}(t)/W_{0x}^{\rm TF}$ and 
$b_z(t)=W_z^{\rm TF}(t)/W_{0z}^{\rm TF}$. 
The initial conditions for the scaling parameters are given by $b_\bot(0)=b_z(0)=1$ and 
$\dot{b}_\bot(0)=\dot{b}_z(0)=0$. 
The initial velocities vanish at $t=0$ because of the equilibrium condition. 
Since we have $\lambda^2\eta + 4\kappa W_{0x}^{\rm TF\,2}\gg 1$ for all experimentally realizable
rotation frequencies $\Omega$, the expansion in $z$-direction 
is much slower than in radial direction. 
Therefore, we may decompose the scaling parameter $b_z=1+\epsilon$ and consider $\epsilon$ as a small time-dependent 
quantity.
In zeroth order in $\epsilon$ we obtain from (\ref{EM5}) for the radial scaling parameter
\begin{eqnarray}
\label{EM6}
b_\bot&=&\frac{\omega_z}{\sqrt{2}\,\Omega}\, \left[\Omega^2/\omega_z^2-\lambda^2\eta
-4\kappa W_{0x}^{\rm TF\,2} \right.\nonumber\\[2mm]
&&\hspace{-1.5cm}\left. +\left(\Omega^2/\omega_z^2+\lambda^2\eta+4\kappa W_{0x}^{\rm TF\,2} \right) 
\cosh{(2\Omega/\omega_z t)}\right]^{1/2}\,.
\end{eqnarray}
With this solution we determine the scaling parameter for the expansion 
in $z$-direction by integrating $\ddot \epsilon=1/b_\bot^2$ either numerically or analytically
in the limit of slow and fast rotation 
frequencies. With this we approximately obtain for $\Omega\ll\omega_z$ 
\begin{eqnarray}
\label{EM7a}
\hspace*{-0.5cm} b_z&\approx& 
1+\frac{t\arctan{\left(\sqrt{\lambda^2\eta+4\kappa W_{0x}^{\rm TF\,2}}t\right)}}{\sqrt{\lambda^2\eta
+4\kappa W_{0x}^{\rm TF\,2}}} 
\nonumber\\
\hspace*{-0.5cm} 
&&\hspace*{-0.8cm} -\frac{ \ln{ \left[1+(\lambda^2\eta+4\kappa W_{0x}^{\rm TF\,2})t^2\right]} }{2\lambda^2\eta
+8\kappa W_{0x}^{\rm TF\,2}}+ \frac{ \Omega t/ 
\omega_z }{ \lambda^2\eta+4\kappa W_{0x}^{\rm TF\,2}}
\end{eqnarray}
and, correspondingly, for $\Omega\approx\omega_z$ 
\begin{equation}
\label{EM7b}
b_z \approx 1+\ln{\left(\cosh{\frac{\Omega t}{\omega_z}}\right)} \,.
\end{equation}
Note the time of flight of the Paris experiment \cite{Bretin,Stock06} lasts 18 ms.
Figure \ref{Ex1} shows that the anharmonicity $\kappa$  has only a marginal effect
on the aspect ratio of the widths 
$W_{x}^{\rm TF} / W_{z}^{\rm TF}= b_x W_{0x}^{\rm TF}/ b_z W_{0z}^{\rm TF}$ 
for a non-rotating condensate ($\Omega =0$). However, this changes drastically once the
condensate is rotating. From Figure \ref{Ex2} we read off that the effect of the trap anharmonicity
upon the aspect ratio is significantly enhanced in the fast rotating regime.
Therefore, this effect must be taken into account for any analysis of free expansion data for large rotation
speeds.
\section{Conclusions}
In this paper we have analyzed dynamical properties of a rotating BEC for the specific parameters
of the Paris experiment \cite{Bretin,Stock06}. With this we have provided useful theoretical results
for future BEC experiments in such anharmonic traps.\\
At first, we have determined how the stationary TF density profile 
of the condensate varies with increasing rotation frequency $\Omega$.
In the overcritical rotation regime we have found that the density in the center is reduced. 
Thus, the rotation frequency has a theoretical upper bound of $\Omega_c\approx1.07\cdot\omega_{\bot}$,
which qualitatively agrees with the highest experimentally realized rotation speed of
$\Omega_{\rm max}\approx 1.04\cdot\omega_\bot$ \cite{Bretin,Stock06}. This quantitative agreement 
is astonishing insofar, as
we have neglected the presence of vortices in our TF analysis.\\
Furthermore, we have used a variational approach to find the frequencies of
the low-energy excitations which strongly depend  on the rotation frequency $\Omega$. 
The anharmonicity influences the eigenmodes in a characteristic way as it enters the equilibrium 
state of the condensate. \\ 
Finally, we have determined the time dependence of the aspect ratio of the condensate widths for the expansion out 
of the trap which is an important quantity for analyzing the experimental data. 
The rotation increases significantly 
the velocity of the expansion in the $xy$-plane. This effect could be used to 
determine the rotation speed in the experiment. \\[0.3cm]
\section{Acknowledgement}
We acknowledge support from the DFG Priority Program SPP 1116 {\it Interaction in Ultracold Gases of 
Atoms and Molecules}. 
%
 
%
\end{document}